\documentclass[showpacs,showkeys,11pt,
preprint,preprintnumbers,nofootinbib,
groupedaddress,superscriptaddress,amsmath,amssymb]{revtex4}
\usepackage{amsfonts}
\usepackage{amsmath}
\usepackage{graphics}
\usepackage{epsfig}
\usepackage{url}
\usepackage{multirow}
\usepackage{feynmp}
\usepackage{xcolor}

\newcommand {\be}{\begin{equation}}
\newcommand {\ee}{\end{equation}}
\newcommand {\ba}{\begin{eqnarray}}
\newcommand {\ea}{\end{eqnarray}}

\begin{document}
\title{Observability of 2HDM Neutral Higgs Bosons in fully hadronic decay at future linear collider}


\author{Nadia Kausar}
\email{nkausar430@gmail.com}
\author{Ijaz Ahmed}
\email{ijaz.ahmed@riphah.edu.pk}
\affiliation{Riphah International University, Islamabad}
\author{Ather M. W.}
\email{mohsan@nutech.edu.pk}
\affiliation{National University of Technology, Islamabad}


\begin{abstract}

The study aims to investigate the observability of pseudoscalar Higgs boson $A$ and neutral heavy CP even Higgs boson $H$,  at different benchmark points, in the framework of  type-I 2HDM. The study is done for $e^{-}e^{+}$ collisions at  $\sqrt{s}$  = 1000 GeV centre of mass energy (c.o.m.) a possible scenario in future lepton collider. 
The associated production of $A$ and $H$ in $e+ e-$ collisions are investigated in fully hadronic final state in two different channels. The first one is $AH \to ZHH \to j\bar{j} b \bar{b}b\bar{b}$ while the other one is $AH \to b \bar{b}b\bar{b}$. The observability of  neutral heavy Higgs and pseudoscalar Higgs signal  is possible within the parameter space $ (\tan \beta, m_{A}) $ which satisfies all  experimental and theoretical constraints. The CP odd and CP even Higgs bosons in all scenarios are observable when signal exceeds $ 5\sigma $, which is the final extracted value of signal significance. The signal significance is calculated at different integrated luminosities. It is concluded from current analysis that fully hadronic channel is promising for search and measurement of the neutral Higgs Bosons in 2HDM.

\end{abstract}
\maketitle

\section{Introduction}
\begin{figure}[h!]
\centering
\includegraphics[width=0.6\textwidth]{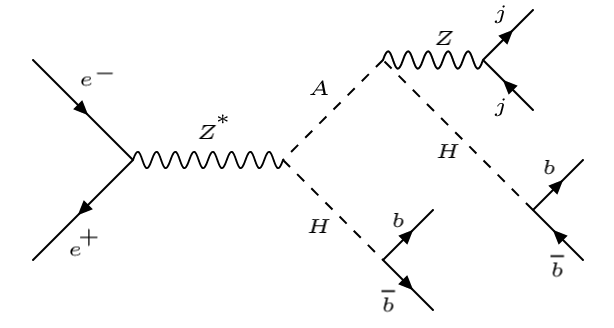}
\caption{\label{fig:diagram1} Feynman diagram for signal process $e^{-}e^{+} \rightarrow AH \rightarrow ZHH \rightarrow jjb \bar{b}b\bar{b}$.}
\end{figure}

After the decades of continuous searches, the Standard Model (SM) Higgs boson was discovered by the CMS and ATLAS experiments at the Large Hadron Collider (LHC) in 2012. This proved to be the corner stone in a very successful theory which passed all of the experimental tests. Yet there are a number of reasons which lead us to believe that this is at best an incomplete theory. This opens a way for physics that is beyond the standard model BSM. The BSM comprises of many theories, but to establish any of these, the experimental evidence is required.  The Two Higgs Doublet Model 2HDM is a model that provides a simple extension to the standard model and features in a number of BSM theories including the Minimal Supersymmetric Standard Model MSSM. There are eight degrees of freedom in two doublet model out of which three degrees of freedom are eaten up by the electroweak bosons due to electro-weak symmetry breaking. The remaining five degrees of freedom lead to five physical Higgs bosons that are 2 CP $2$ CP-even neutral scalar Higgs bosons $h$ , $H$ , a CP-odd pseudoscalar neutral Higgs boson $A$ and a pair of charged Higgs bosons $H^{\pm}$. The existence of these Higgs bosons are  to be verified through experimental measurements of production cross sections and branching ratios. In  July 2017,  ATLAS  published  the results for collaboration on the trending search of decaying neutral Higgs collaboration into two tau leptons. The tau leptons are individually interesting to search due to the existence of strong coupling of A/H and fermions of down type for the specific value of MSSM provided parameter space. These will certainly boost up the probability of A/H production in accordance with b-quarks and provides a higher cross-section. Future  lepton collider such as $e^{-}e^{+}$  will play an important role for the detection of Higgs boson. The precise attention is given to  search the Higgs boson. The main focus of this study is to investigate the  production of  neutral Higgs $ A $ in association to $ H $  at electron positron collider. The assumed framework for the study is Type 1 2HDM. Several benchmark points are assumed with different mass hypotheses. Two decay channels are investigated in this study. In the first one, the pseudoscalar Higgs $A$ decays to $Z$ boson along with neutral CP even heavy Higgs boson $H$, the Z boson then decays hadronically to quarks while the H boson decays to $b \bar{b}$ leading to two light jets and four b jets in final state ($j\bar{j} b \bar{b}b\bar{b}$). The Feynman diagram for first process is given in Figure~\ref{fig:diagram1}.

\begin{figure}[h!]
\centering
\includegraphics[width=0.6\textwidth]{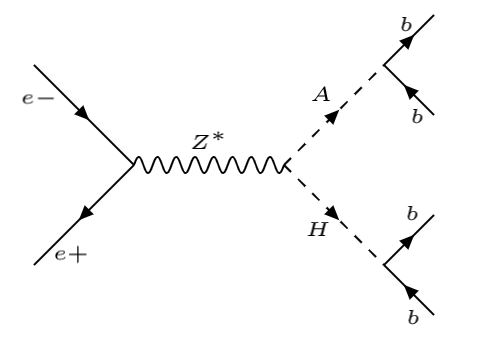}
\caption{\label{fig:diagramF} Feynman diagram for signal process $e^{-}e^{+} \rightarrow AH  \rightarrow b \bar{b}b\bar{b}$.}
\end{figure}

In the second decay channel both A and H decay to b-quarks leading to four b jets in final state ($b \bar{b}b\bar{b}$). The Feynman diagram of this process is given in Figure~\ref{fig:diagramF}. 
\section{Two Higgs Doublet Model 2HDM}
The 2HDM is extension of the  SM  which is featured in many BSM theories. This model is proficient in  solving some of the problems in SM while still maintaining the good agreement between the SM and experiments. The 2HDM offers the most simple extension of the SM where the Higgs sector is extended by including an additional doublet in the theory. The Supersymmetric theories show that the scalars are associated to chiral multiplets and opposite chirality is found in their complex conjugates. The most attractive feature of 2HDM that it smoothens the way for initiating new possibilities for explicit or for the automatic CP violation ~\cite{fk}. Each class of 2HDM gives interesting environment and unique phenomenology. The 2HDM is divided into  four types depending on the coupling of fermions with the doublets as shown in Table~\ref{tab:1}.
\begin{table}[h]
	\centering
\begin{tabular}{|l|c|c|c|c|}
\hline
Types of Model& Description&Up quarks   & Down quarks  &  charged leptons \\
\hline
Type I & Fermiophobic&$\Phi_{2}$ & $\Phi_{2}$ &  $\Phi_{2}$ \\
\hline
Type II &MSSM like &$\Phi_{2}$ & $\Phi_{1}$ & $\Phi_{1}$ \\
\hline
X&Lepton-specific & $\Phi_{2}$ & $\Phi_{2}$ & $\Phi_{1}$ \\
\hline
Y&Flipped & $\Phi_{2}$ & $\Phi_{1}$ & $\Phi_{2}$ \\
\hline
\end{tabular}
\caption{\label{tab:1}Different types of 2HDM on the  basis of coupling of Fermions with charged leptons }
\end{table}

General scalar of 2HDM is the  most usual scalar potential of 14 parameters which can function as charge parity (CP) conserving, parity-violating and charge violating minima. The common scalar potential expression is assumed for two doublets $ \phi_{1} $ and $\phi_{2}$ ~\cite{gf} with hypercharge +1  and is given in equation \eqref{eq:A}
\begin{equation}
\label{eq:A}
\begin{aligned}
V_{2HDM}=m_{11}^{2}\phi_{1}^{\dagger}\phi_{1}+m_{22}^{2}\phi_{2}^{\dagger}\phi_{2}
-m_{12}^{2}(\phi _{1}^{\dagger}\phi _{2}+\phi _{2}^{\dagger}\phi _{1})+\frac{1}{2}\lambda_{1}(\phi _{1}^{\dagger}\phi_{1})^{2}+\frac{1}{2}\lambda_{2}(\phi_{2}^{\dagger}\phi_{2})^{2}+\lambda_{3}(\phi _{1}^{\dagger}\phi_{1}) (\phi_{2}^{\dagger}\phi _{2})+\\
\lambda_{4}(\phi_{1}^{\dagger}\phi_{2})(\phi _{2}^{\dagger}\phi_{1})+\frac{1}{2}\lambda_{5}[(\phi_{1}^{\dagger}\phi_{2})+(\phi_{2}^{\dagger}\phi _{1})]
\end{aligned}
\end{equation}
This potential contains all real parameters with two assumed $SU(2)$ doublets in 2HDM. $v_{1}$ and $ v_{2} $ are the values of vacuum expectation of two doublets i.e $\phi_{1}$ and $\phi_{2}$. The doublet expanded by making known to eight read field $ w_{i}^{\pm} $, $\rho_{i}$, $ \eta_{i} (i=1,2......)$ all over the place of these minima, is given in equation \eqref{eq:b} 
\begin{equation}
\label{eq:b}
\langle\phi_{1}\rangle=\frac{1}{\sqrt{2}}
\begin{pmatrix}
0 \\ 
v_{1}
\end{pmatrix}, 
\langle\phi_{2}\rangle=\frac{1}{\sqrt{2}}
\begin{pmatrix}
0 \\ 
v_{2}
\end{pmatrix}
 \end{equation}
 \begin{equation}
\label{eq:z}
\langle\phi_{1}\rangle=
\begin{pmatrix}
w_{1}^{\dagger} \\ 
\dfrac{v_{1}+\rho_{1}+\iota\eta_{1}}{\sqrt{2}}
\end{pmatrix}, 
\langle\phi_{2}\rangle=
\begin{pmatrix}
w_{1}^{\dagger} \\ 
\dfrac{v_{2}+\rho_{2}+\iota\eta_{2}}{\sqrt{2}}
\end{pmatrix}
 \end{equation}
 By implementing the two minimization conditions of 2HDM, the terms $ m_{11}^{2} $ and $ m_{22}^{2} $ can be eliminated  in the favour of pseudo scalar inputs. By the use of conditions  seven real independent parameters ~$ \tan\beta=\dfrac{v_{2}}{v_{1}}, m_{11}^{2},\lambda_{1},\lambda_{2},\lambda_{3},\lambda_{4},\lambda_{5} $ are obtained.\\
 To be more convenient, the  other parameters would be $m_{h}, m_{H}, m_{A}, m_{H^{\pm}}, \alpha, \tan\beta, m_{12}^{2}$. Where $  \alpha $ is mixing angle which rotates non-physical states  ($\phi_{1}$ and $\phi_{2}$ ) to the physical states ($ h $ and $ H $). For Yukawa Lagrangian, the procedure as in SM cannot be followed and if both of the Higgs doublets couples to SM fermions, this would be flavor violating. To overcome this problem, the discrete symmetry $ Z_{2} $ is imposed.\\
A discrete $ Z_{2}$ symmetry ~\cite{gs} was forced on the 2HDM potential to avoid FCNCs ~\cite{ga}  but the potential still contains a term that clearly breaks this symmetry. If $ m_{12}^{2} $ is not disappearing, the potential is not invariant under the transformation $ \phi_{1}\rightarrow\phi_{2} $ but this form of symmetry breaking is only soft because $ m $ has mass dimension.
The constraints are applied to 2HDM parameters by some of the theoretical considerations like perturbativity, vacuum stability, perturbative unitarity. Consistency is checked at the confidence level of $95\%$.
\section{Signal Process}
The  observability of  neutral Higgs bosons  is investigated in two final states. In the first one we have signal chain process  $e^{-}e^{+} \rightarrow AH \rightarrow ZHH \rightarrow j\bar{j} b \bar{b}b\bar{b}$  and  in second final state the chain process is  $e^{-}e^{+} \rightarrow AH \rightarrow b \bar{b}b\bar{b}$ within framework of Type 1 2HDM in the low $tan\beta$  regime and  enhancement of $A$ and $H $ is achieved. The  $Z$ boson goes into hadronic decay as a di-jet pairs $j\bar{j}$, $Z \rightarrow j\bar{j}$. The enhancement of the Higgs boson decay mode $H \rightarrow b\bar{b}$ is due to $cot\beta$ factor. The  several benchmark points with different mass   hypothesis are shown in Table~\ref{tab:signalcuts1}. At linear collider, the initial collision is assumed to take place at the centre of mass energy $\sqrt{s} =  1000$ GeV and integrated luminosity is assumed to be 500 $ fb^{-1} $. The range for  mass of the  CP-even Higgs boson $ H $  is assumed to be from $150 - 300$ GeV  and for CP-odd Higgs boson $A $, the assumed  range  is $250 - 400 $ GeV with the mass splitting of $50-100 $ in all the scenarios. The value of $tan\beta$ is set to $10$ for all scenarios which results in the enhancement of H decay. To satisfy theoretical requirement of potential stability \cite{sp} , there is a range of $m_{12}^2$ parameter for each scenario. By using 2HDMC-1.7.0 \cite{con}, perturbativity and unitarity constraints are checked.\\

  Branching ratios (BR) are achieved by using 2HDMC-1.7.0 ~\cite{con}. B-tagging and jet reconstruction algorithms give rise in the uncertainties which perturb the final results due to more errors in the hadronic $Z$ boson decay. The $Z$ boson decay provides a simple and clean signature at linear collider. The branching ratio of hadronic  decay of  $Z$ boson  is BR$(Z \rightarrow q\bar{q})  \cong 0.69$  and the BR for  $H$ decay  is BR $(H \rightarrow b\bar{b}) \cong  0.71$. In $AH \to ZHH$ process, the $Z$ and $H$ Higgs boson are reconstructed by recombination of two light-jets and two b-jets respectively.  The pseudoscalar $A$ Higgs boson is then reconstructed by the reconstructed $Z$ and $H$ bosons. In the second process, both $ A $ and $ H $ are reconstructed by recombination of  b-jet pairs.\\
  The  main SM background processes  which are taken into account for this analysis, relevant to signal are pair production of di-vector boson  $W^{\pm}$, $ZZ$, top quark $ t\overline{t} $ and  $Z/\gamma$ production. Cross sections are computed at $\sqrt{s}= 1000$ GeV  by PYTHIA 8.2.10 ~\cite{pythia1} and are given in Table~\ref{tab:signalcuts1}. The cross section for first process is represented by $ \sigma_{1} $ and for second process it is represented by $ \sigma_{2} $
 \begin{table}[h!]
\centering
\begin{tabular}{|c|c|c|c|c||c|c|c|c|c|c|c|c|c|}
\hline
		
		& BP1 & BP2 & BP3&BP4 &  $W^{\pm}$ & $t\overline{t} $& $ZZ$ & $Z/\gamma$ \\
		\hline
		$m_{h}$ & 125 & 125  & 125    &125&&& &  \\
		\hline
		$m_{H}$  & 150  & 200 & 250  &300&&& &\\
		\hline
      	$m_{A}$	& 250 & 300 & 330 &400&&&&\\
		\hline
		$m_{H^\pm}$	& 250 & 300 & 330&400&&&&\\
		\hline
		$m_{12}^{2}$	& 1987-2243 & 3720-3975 &  5948-6203&8671-8925&&&& \\
		\hline 
		$tan\beta$&10&10&10&10&&&&\\
		\hline
		$sin(\beta - \alpha)$ &1&1&1&1&&&& \\
		\hline
		$\sigma1(fb)$ at $ \sqrt{s}=1000 $ GeV &10.16&8.211&6.675&4.213&3192&210.8&69.24&2328\\
		\hline
		$\sigma1(fb)$ at $ \sqrt{s}=1500 $ GeV &5.587&5.137&4.725&3.937&1812&102.6&38.92&1063\\
		\hline
		$\sigma1(fb)$ at $ \sqrt{s}=3000 $ GeV &1.692&1.625&1.603&1.484&687.5&28.94&13.79&278.8\\
		\hline
	$\sigma2(fb)$ at $ \sqrt{s}=1000 $ GeV 	&10.16&8.221&6.686&4.212&3192&211.4&109.3&2874\\
	\hline
	$\sigma2(fb)$ at $ \sqrt{s}=1500 $ GeV 	&5.609&5.090&4.694&4.001&1815&101.8&61.38&1320\\
	\hline
	$\sigma2(fb)$ at $ \sqrt{s}=3000 $ GeV 	&1.681&1.628&1.59&1.48&687.2&28.89& 21.59& 347.6\\
		\hline
		
\end{tabular}
\caption{\label{tab:signalcuts1}  The cross section for signals at different benchmark points  and for SM background processes.}
\end{table}
 The Branching Ratio for decay of  neutral Higgs boson H  into bottom quark pairs  for our assumed  BP points are  calculated by using 2HDMC-1.7.0.
 Branching ratios of heavy  neutral Higgs bosons H and  pseudo scalar A Higgs boson decay  for different benchmark points are tabulated in the Table~\ref{tab:signalcuts3a}  and Table~\ref{tab:signalcuts3b} respectively. 
\begin{table}[h]
\centering
\begin{tabular}{|c|c|c|c|c||c|c|}
\hline
		
		BP& $BR(H\rightarrow b\overline{b})$& $BR(H\rightarrow \tau\tau)$& $BR(H\rightarrow gg)$  \\
		\hline
		$200$ & $ 6.755 \times10^{-1}$ &$7.133\times10^{-2}$&$2.22\times10^{-1} $   \\
		\hline
		$225$  & $4.474\times10^{-1}$& $4.831\times10^{-2}$&$1.957\times10^{-1}$ \\
		\hline
      	$250$	& $4.668\times10^{-2}$&$5.139 \times10^{-3}$&$ 2.681 \times10^{-2}$\\
		\hline
		$275$	&  $ 4.568 \times10^{-3}$ &$5.118 \times10^{-4}$&$ 3.429 \times10^{-3}$\\
		\hline 
		$300$&$ 7.625\times10^{-5}$&$8.679\times10^{-6} $&$ 7.525 \times10^{-5}$\\
		\hline
		$325$ &$1.839\times10^{-5}$&$ 2.124 \times10^{-6}$&$2.446\times10^{-5}$ \\
		\hline
		
\end{tabular}
\caption{\label{tab:signalcuts3a}The  branching ratio of neutral Higgs boson H decay for benchmark points .}
\end{table}
\begin{table}[h]
\centering
\begin{tabular}{|c|c|c|c|c||c|c|}
\hline
		
		BP& $BR(A\rightarrow ZH)  $ & $ BR(A\rightarrow Z\gamma) $&$ BR(A\rightarrow gg) $&$ BR(A\rightarrow Z\mu) $  \\
		\hline
		$200$ & $6.279\times10^{-1}$ & $ 5.590\times10^{-5} $ &$1.6666\times10^{-1}$&$6.777\times10^{-5}$   \\
		\hline
		$225$  & $ 0.9098 \times10^{-1}$ & $ 1.582\times10^{-5} $&$ 3.839 \times10^{-2}$&$ 1.188\times10^{-5} $  \\
		\hline
      	$250$	& $ 0.9896\times10^{-1} $ &$ 4.279 \times10^{-7}$ &$ 8.944 \times10^{-4}$&$2.101\times10^{-7} $\\
		\hline
		$275$	& $ 0.9863\times10^{-1} $ & $8.6\times10^{-8}$ &$ 1.603\times10^{-4} $&$ 2.814 \times10^{-8}$\\
		\hline 
		$300$&$ 0.7406 \times10^{-1}$&$ 4.308 \times10^{-8}$&$7.327 \times10^{-5}$&$9.259 \times10^{-9}$\\
		\hline
		$325$ &$0.5605 \times10^{-1} $&$ 3.303\times10^{-8} $&$ 5.120\times10^{-5} $&$ 3.792\times10^{-9} $ \\
		\hline
		
\end{tabular}
\caption{\label{tab:signalcuts3b}The  branching ratio of Pseudo scalar A decay for benchmark points .}
\end{table}

\section{EVENT GENERATION, SIGNAL SELECTION AND ANALYSIS}
    The parameters of Type-1 are produced in  SLHA (SUSY Les Houches Accord) format  by using 2HDMC-1.7.0 ~\cite{con}. This output file is passed to PYTHIA 8.2.10 ~\cite{pythia1} to generate the  events. The generated particles in each event are then passed to clustering algorithm fastjet-3.3.3 to make jets ~\cite{mm}. In order to record the events data, the interface of HepMC-2.06.06~ \cite{ Hep} is given to Pythia. The output of Pythia is then analyzed and histograms are plotted by using Root- 6.20/04 ~\cite{root}. 

    Different kinematic selection cuts are applied which arises  certain fluctuations in the signal. These cuts define the band of ranges which are invariant quantities, measured in events. It must fulfil the number of several final state particles. These particles are identified in phase of primary reconstruction using “object identification cuts”. Then the kinematic selection cuts are applied to refine the rejection and selection of background events to finalise the results.\\

The first step in event selection is the kinematic cut on jets which omit the soft $p_T$ jets and the ones that are in forward region along the collision beams. For this we apply following cuts on transverse momentum and pseudorapidity of jets.
\begin{equation}
\label{eq:1}
p_{T}^{jet}  > 20 GeV , | \eta_{jets} | <  2.5
\end{equation} 
Once we have jets within the desired kinematic range, we split the reconstructed jets by identifying them as light and b-tagged jets. In order to achieve this, we do a $\Delta R$ matching of the jets with the generated particles which is defined as 
\begin{equation}
\label{eq:2}
\Delta R=\sqrt {(\Delta \eta)^{2}+(\Delta \phi)^{2}}
\end{equation}

We identify the jets which are within $\Delta R < 0.4$ of the b-quarks in the event as b-jets and the ones that are farther away from b-quarks as light jets. Once we have identified the jets, we apply the multiplicity cut on the jet. For $j\bar{j} b \bar{b}b\bar{b}$ channel we require the event to have at least two light jets and at least four b-jets while for $b \bar{b}b\bar{b}$ we require at least 4 b-tagged jets.
   
In $AH \to ZHH \to j\bar{j} b \bar{b}b\bar{b}$ analysis, we use the selected light-jets (b-jets) for find a combination that minimise the $\chi^2$ defined as follows.

\begin{equation}
\chi^2 = \left(\frac{m_{jj} - m_Z}{\sigma_{m_Z}}\right)^2 + \left(\frac{m_{bb,1} - m_H}{\sigma_{m_H}}\right)^2 + \left(\frac{m_{bb,2} - m_H}{\sigma_{m_H}}\right)^2
\end{equation}  

where $m_{jj,bb}$ are the dijet mass, $m_Z$ is the mass of $Z$ boson, and $m_H$ is the mass of the heavy Higgs boson according to the BP taken, and the $\sigma_{m_{Z,H}}$ are the widths of the respective mass distributions. The cut of $\chi^2 < 10$ is applied to select only events with good reconstructed $Z$ and $H$ bosons. The pseudoscalar Higgs boson $A$ is then reconstructed using the combination of $Z$ and $H$ which gives mass nearest to $A$ nominal mass according to the BP.

The same treatment is given to $AH \to b \bar{b}b\bar{b}$ analysis but in this case, $\chi^2$ is defined as 

\begin{equation}
\chi^2 = \left(\frac{m_{bb,1} - m_A}{\sigma_{m_A}}\right)^2 + \left(\frac{m_{bb,2} - m_H}{\sigma_{m_H}}\right)^2
\end{equation}

After applying different selections cuts,  The efficiencies are calculated for different benchmark points and are given in Table~\ref{tab:signalcuts2} for first process  and Table~\ref{tab:signalcuts2h} for second process  respectively.
 
\begin{table}[h]
\centering
\begin{tabular}{|c|c|c|c|c||c|c|c|c|}
\hline
		Cuts
		& BP 1  &  BP 2  & BP 3 &  BP 4	\\
		\hline
		 Two light jets & 0.4078&0.5122&0.5072&0.6249\\
		\hline
		  Four b-jets &   0.5945&0.6364	&0.6270&0.6496\\
		 \hline
        	$ \chi^{2} $	  & 0.6356	&0.5271&0.2910&0.3408\\ 
		\hline 
		Total Efficiency  &0.1541&	0.1718&0.09254&0.1384\\
		
		\hline
		$\sigma$ x BR[$fb$] &4.481 &2.886 &1.715 &0.580\\
		\hline
\end{tabular}
\caption{\label{tab:signalcuts2} The efficiencies for different selection cuts at different mass hypothesis in first process.}

\end{table}
\begin{table}[h]
\centering
\begin{tabular}{|c|c|c|c|c||c|c|c|c|}
\hline
		Cuts
		& BP 1  &  BP 2  & BP 3& BP4   \\
		
		\hline
		  Four b-jets &0.7161&0.7221&0.7167&0.7059  \\
		  \hline
		  $\chi^{2} $&0.7764&0.6766&0.6036&0.5053  \\
		  \hline
		  $A\chi^{2}$ &0.8118&0.8013&0.8024&0.8207  \\
		\hline 
		Total Efficiency&0.4514&0.3915&0.3471&0.29\\
		\hline
		$\sigma$ x BR[$fb$] & 4.09042$\times10^{-3}$ & 2.94312$\times10^{-2}$  & 6.79966$\times10^{-2}$ &  4.74 $\times10^{-4}$  \\
		\hline
\end{tabular}
\caption{\label{tab:signalcuts2h} The efficiencies for different selection cuts at different mass hypothesis  in second process.  }

\end{table}
\section{RESULT AND DISCUSSION}
\begin{figure}[h!]
\centering
\includegraphics[width=.78\textwidth]{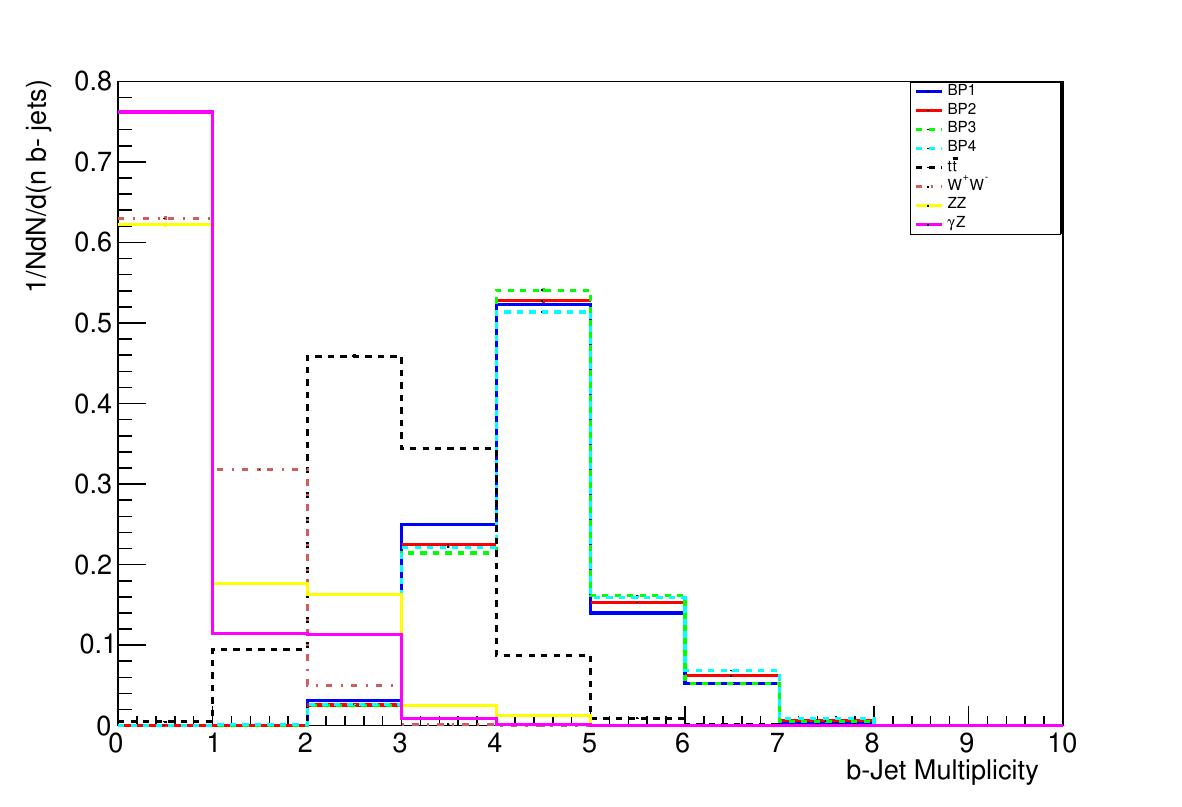}

\caption{\label{fig:diagram5} The distributions of b-jet multiplicity  corresponding to different background and signal processes at CMS energy of 1000 GeV.}
\end{figure}

\begin{figure}[h!]
\centering
\includegraphics[width=.78\textwidth]{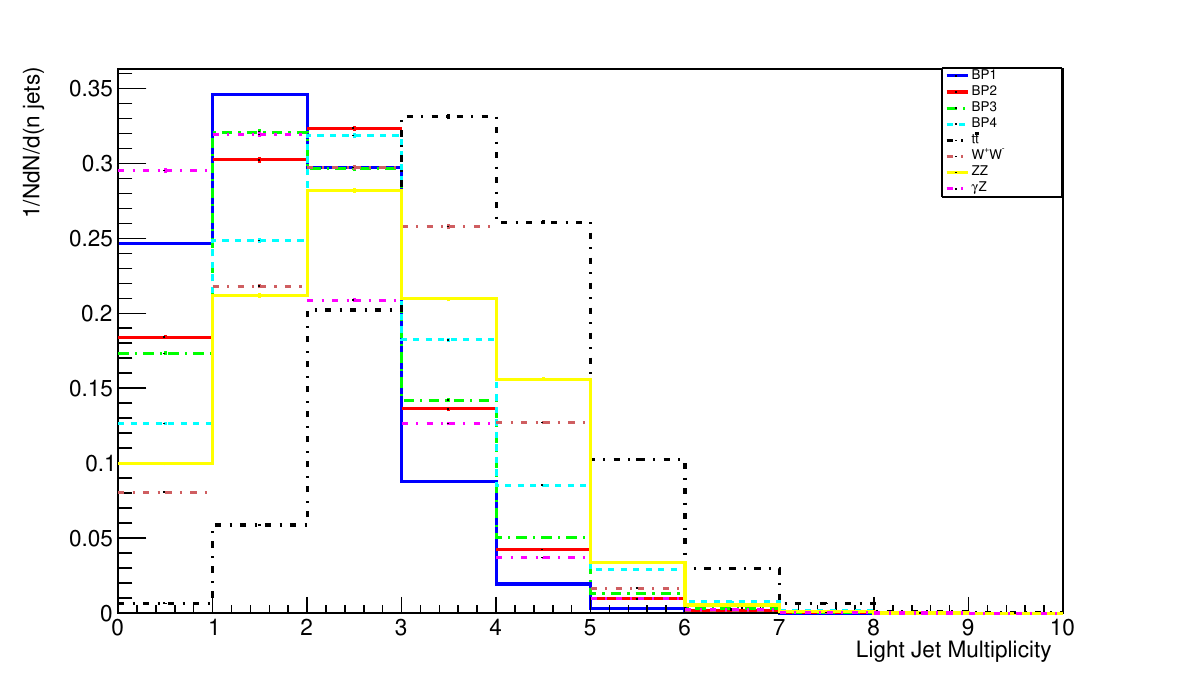}

\caption{\label{fig:diagram5a} The distributions of light jet multiplicity  corresponding to different background and signal processes  at CMS energy of 1000 GeV.}
\end{figure}

\begin{figure}[h!]
\centering
\includegraphics[width=.78\textwidth]{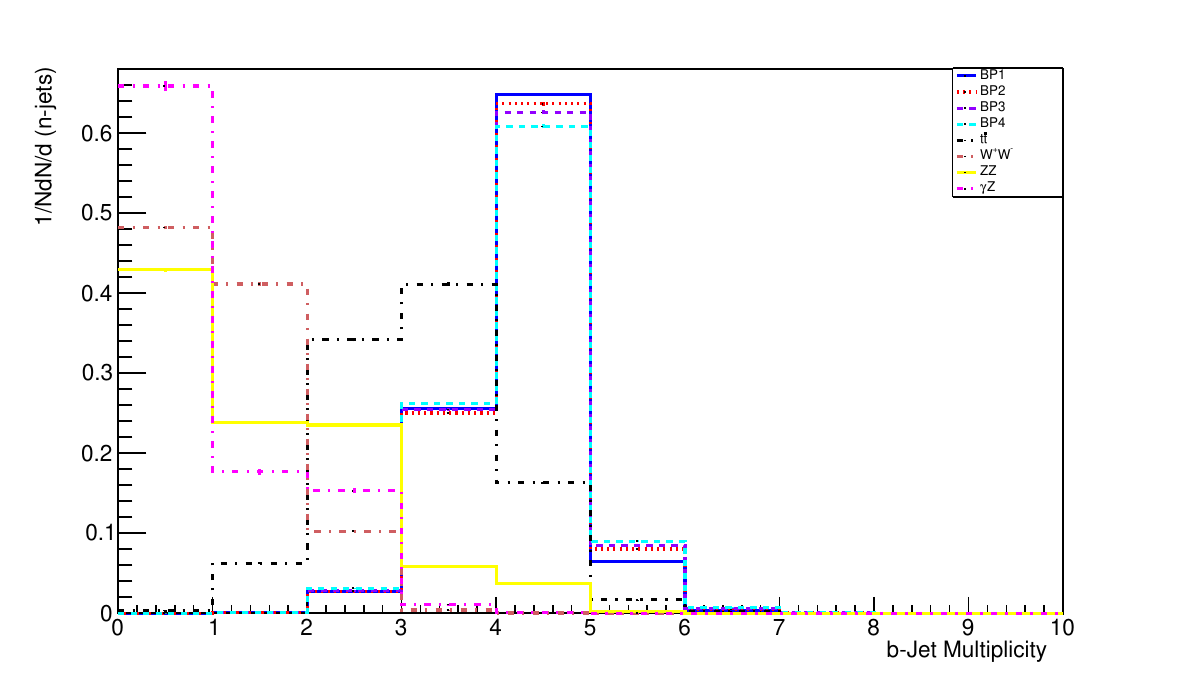}

\caption{\label{fig:diagram5a1} The distributions of b-jet multiplicity for corresponding to different background and signal processes  at CMS energy of 1000 GeV.}
\end{figure}

 In the finalized results, topology of our assumed  first signal process contains four b-jets and two light jets. The distribution of b-jet multiplicity for process having 4 bjets and 2 light jets  is depicted  in Figure~\ref{fig:diagram5} . 
 More than 60 percent of the jets are passed through the jets kinematic cut.
The distribution of light jet multiplicity  for this process   is shown in Figure~\ref{fig:diagram5a} for both background and signal events. 
The other process contains 4 b-jets. The distribution of b-jet multiplicity for this  process   is shown in Figure~\ref{fig:diagram5a1} . 

\begin{figure}[h!]
\centering
\includegraphics[width=.78\textwidth]{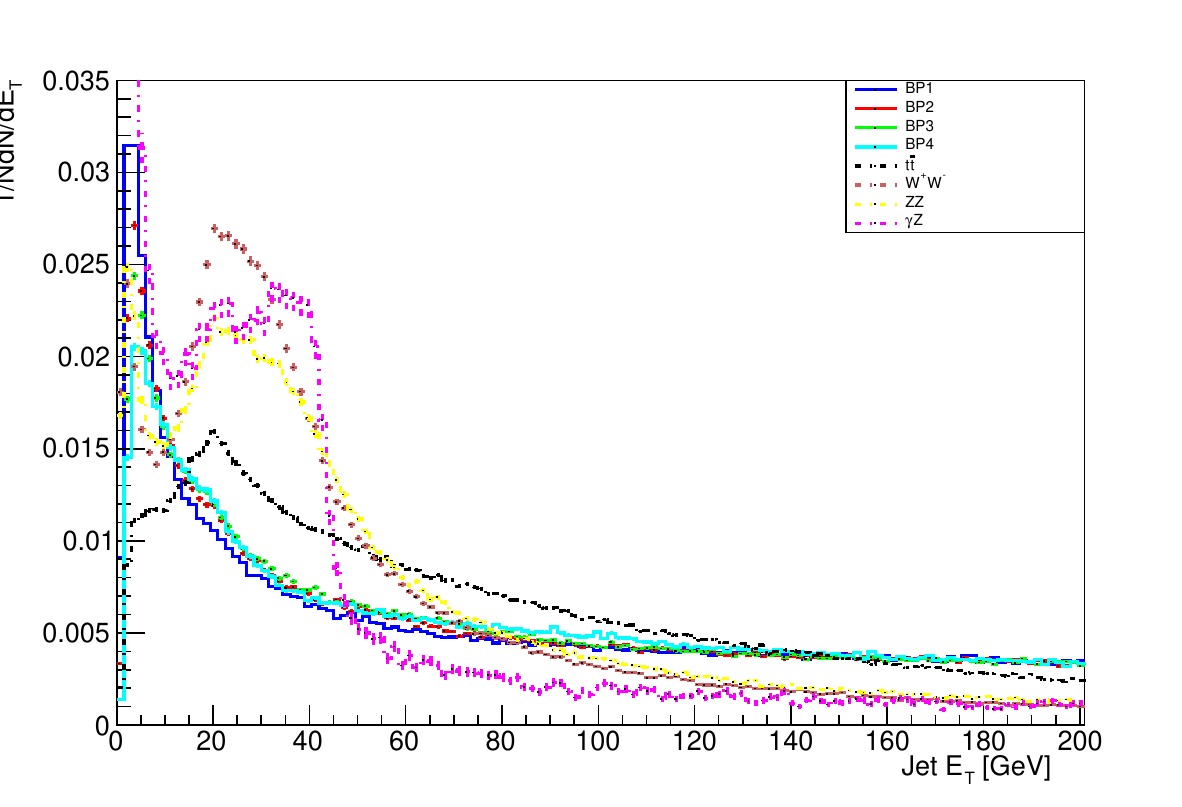}

\caption{\label{fig:diagram24} Transverse energy of jets.}
\end{figure}

\begin{figure}[tb]
\centering
\includegraphics[width=.78\textwidth]{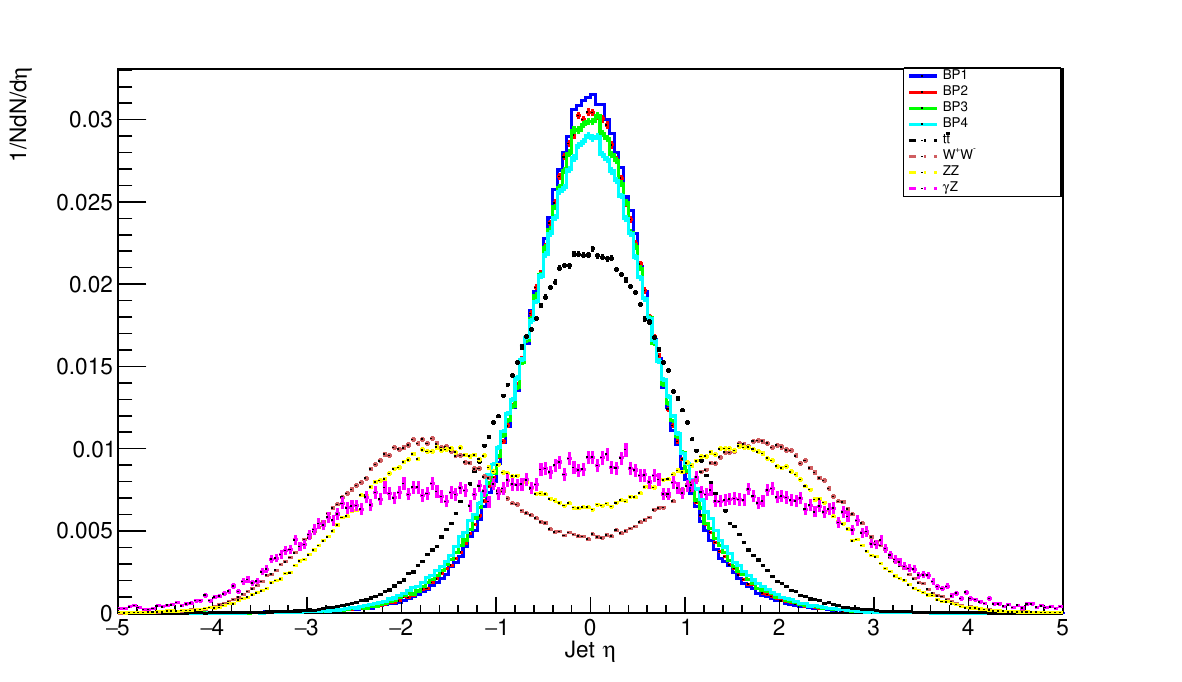}

\caption{\label{fig:diagram2} The pseudorapidity of jets.}
\end{figure}

The distribution of b-jets slightly depends on neutral Higgs boson mass in the assumed signal event. The production of b-jet is suppressed kinematically in the production of  Higgs boson, the availability of phase space is smaller due to which it decays to bottom quarks. 
 Transverse energy  $E_{T}$ of jets for second process  is shown in figure~\ref{fig:diagram24}.
 The pseudorapidity of jets for  this process   is shown in figure~\ref{fig:diagram2}.

\begin{figure}[h!]
\centering
\includegraphics[width=.78\textwidth]{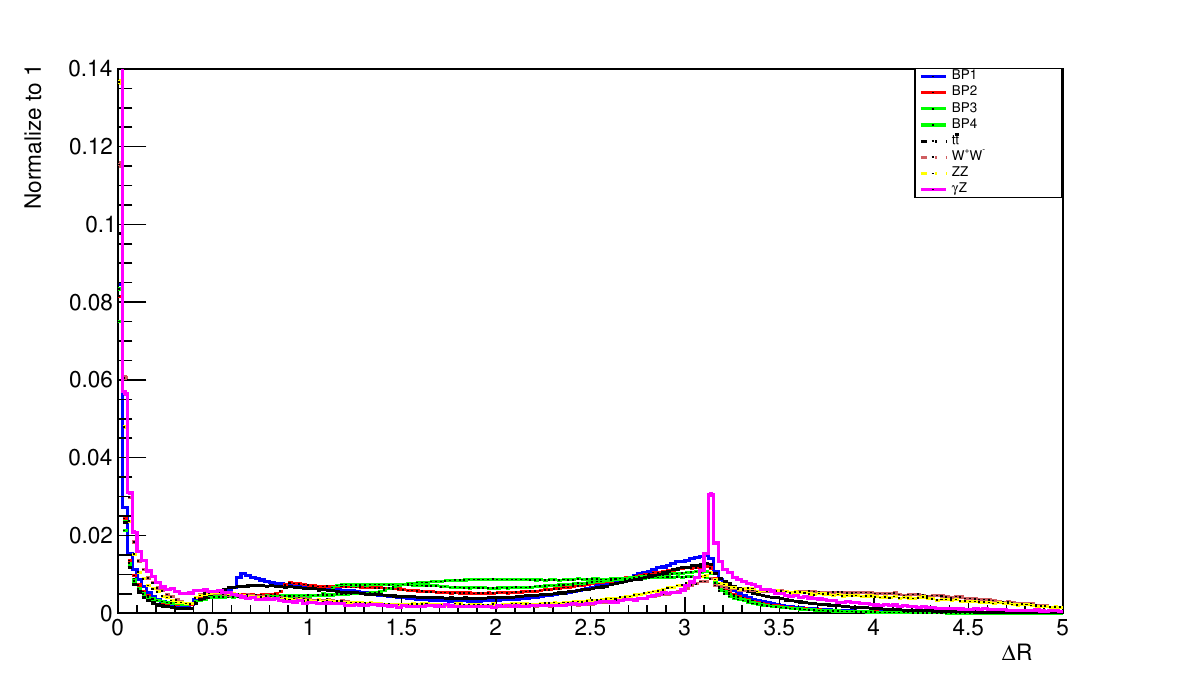}

\caption{\label{fig:diagram6}The distribution of  $ \Delta$ R( jets, quarks) variables used to tag bjet for all signal and background processes.}
\end{figure}

After extracting out the data of $\Delta R$,  the profiling process of $\Delta R$ is discussed. By the analysis of plot of $\Delta R$ shown in Figure~\ref{fig:diagram6},  the b-jets can easily be  identified by finding the minima of the plot. 
For all the distribution profiles, the implemented integrated luminosity is $500 fb^{-1}$ . The cross-sections of the signal are collected by multiplying the branching ratios with corresponding total cross-sections and  are given in Table ~\ref{tab:signalcuts2} for first process and in Table ~\ref{tab:signalcuts2h} for second process.

\subsection{Mass reconstruction}
 The invariant mass of neutral scalar Higgs bosons is reconstructed. In this work, the generation of  signal and background processes is involved which display natural interference and selection techniques where different mass speculations are displayed for  Higgs invariant mass remaking.\\
The process of mass reconstruction of Higgs bosons is considered as  important to attain the reliable separation between the main assumed signal and the background processes. Peak of the signal resonance would be produced by the proper mass variable. By this process, large signal will be produced over the background ratio. The pair of b-jets comes out from the heavy Higgs Boson in the signal events. Due to that fact, the invariant mass of this pair should be lesser and lie within mass casement adjusted by neutral Higgs mass. Mass of the Higgs Boson can be  calculated by the conventional formula given as
\begin{equation}
\label{eq:9}
m_{H}=\sqrt {E^{2}-p_{x}^{2}-p_{y}^{2}-p_{z}^{2}}
\end{equation}
\section{Chi-square method}
\begin{figure}[tb]
\centering
\includegraphics[width=.78\textwidth]{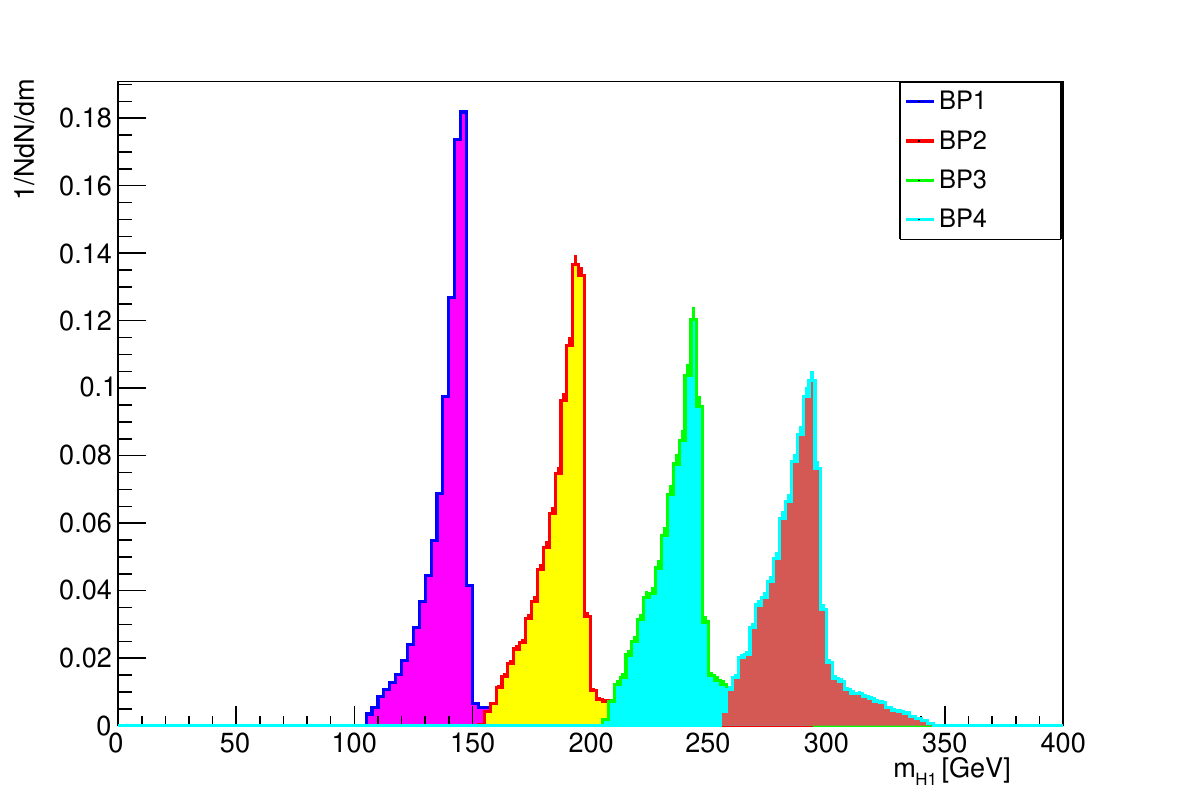}

\caption{\label{fig:diagram8}The reconstructed mass of neutral Higgs $ H_{1} $ at different mass hypothesis for first process .}
\end{figure}

\begin{figure}[h!]
\centering
\includegraphics[width=.78\textwidth]{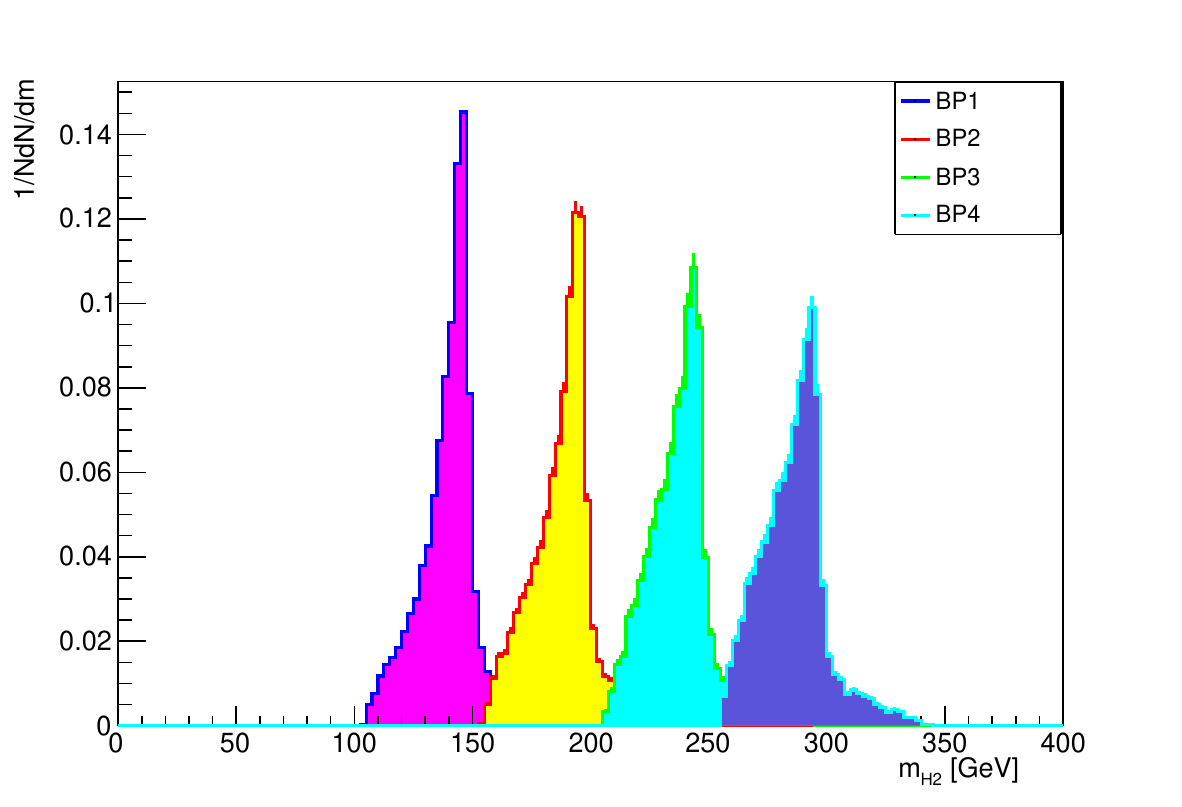}

\caption{\label{fig:diagram8a}The reconstructed mass of neutral Higgs $ H_{2} $ at different mass hypothesis for first process .}
\end{figure}

Chi-square test is based on comparing observed and expected values. This test is designed to determine whether a difference between actual and predicted data is due to chance or due to relationship between the variables under consideration. As a result, Chi-square test is a fantastic tool for analyzing and understanding the relationships between our categorical data. The Chi-square method is used in this study to compare the reconstructed and actual values of Higgs boson mass. To determine the best estimate of Higgs-boson mass, I minimized the appropriately weighted sum of squared differences between observed and predicted values, known as "Chi-square".
The general formula for Higgs boson is given in equation ~\eqref{eq:9z}

\begin{equation}
\label{eq:9z}
\chi^{2}=\left(\dfrac{m_{Rec}-m_{Gen}}{\sigma_{dijet}}\right) ^{2}
\end{equation}
The reconstructed invariant mass of the Higgs boson is represented by "$m_{rec}$", and the actual value of the Higgs boson mass is represented by "$m_{Gen}$". Then, for each possible combination, the sum of squared differences between observed and predicted Higgs-boson masses is calculated. Then, light jet and bjet pairings that meet the following conditions are chosen.
\begin{center}

$\chi^{2}_{min}>20$
\end{center}
Then, using this true light jet pair, the Higgs-boson mass is reconstructed. The reconstructed invariant mass of  Higgs-boson from b jet pairs and light-jet pairs for all signal along with SM backgrounds are shown in  figures.
 The figures Figure~\ref{fig:diagram8} and Figure~\ref{fig:diagram8a}  represents the reconstruction of  neutral scalar Higgs for first process.


\begin{figure}[h!]
\centering
\includegraphics[width=.78\textwidth]{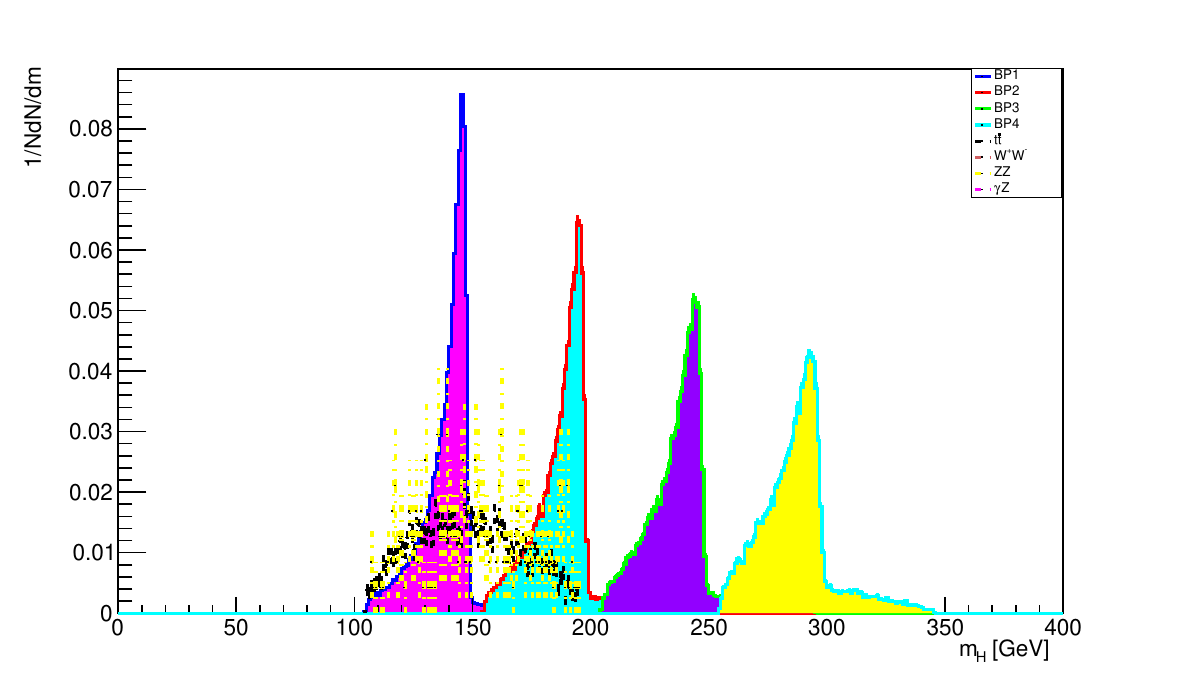}

\caption{\label{fig:diagram7}The reconstructed mass of neutral Higgs at different mass hypothesis  for second process.}
\end{figure}

The reconstruction of neutral Higgs boson  is plotted in Figure~\ref{fig:diagram7}  for second  process.
The mass peak of neutral scalar Higgs can be clearly seen in the Figure~\ref{fig:diagram7}  which suppresses the significant background processes. From Figure~\ref{fig:diagram8} , Figure~\ref{fig:diagram8a} and  Figure~\ref{fig:diagram7}  it can be examined that the attained data fall nearly to the input mass which is only possible due to the testing of different mass hypothesis of neutral scalar Higgs in simulation.\\
\subsection{Mass Reconstruction of Pseudo scalar Higgs}
\begin{figure}[h!]
\centering
\includegraphics[width=.78\textwidth]{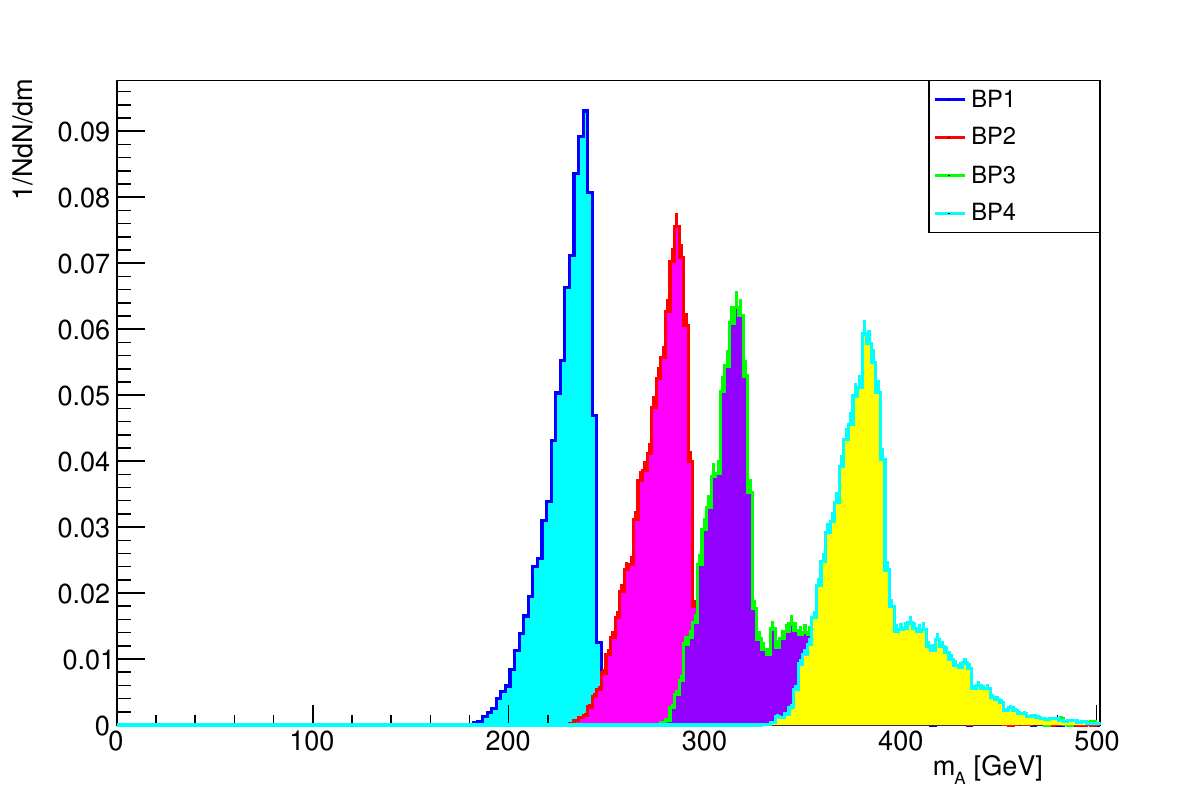}

\caption{\label{fig:diagram9} The mass reconstruction of pseudo scalar Higgs A for first process .}
\end{figure}

\begin{figure}[h!]
\centering
\includegraphics[width=.78\textwidth]{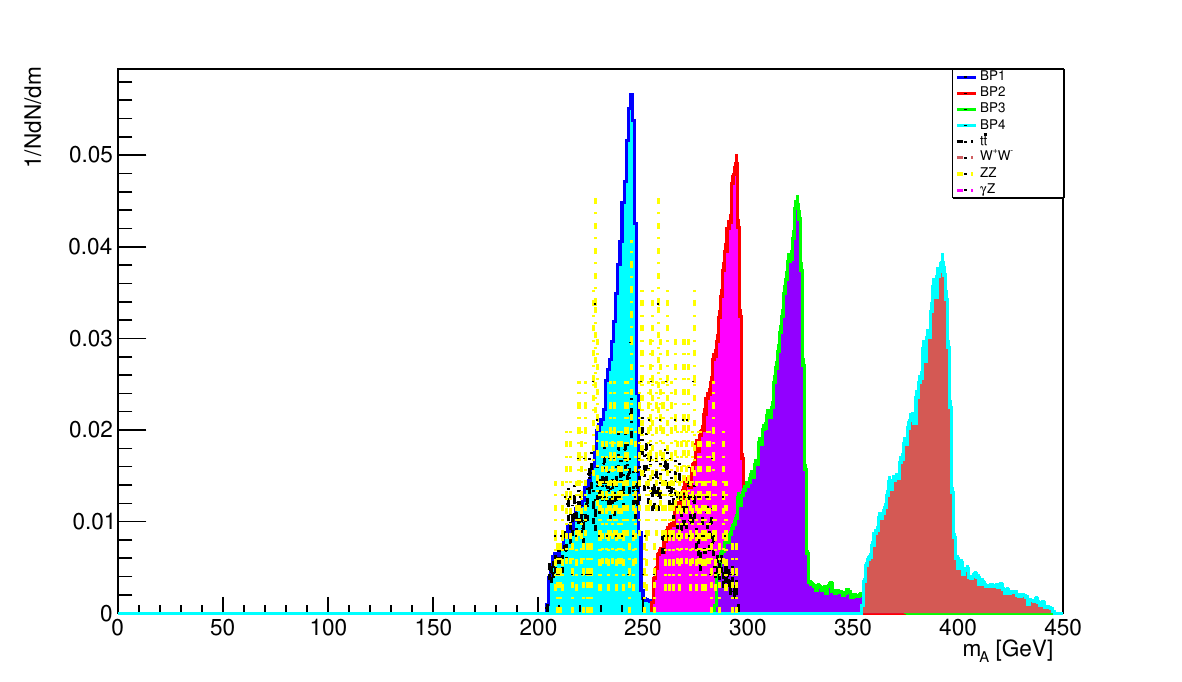}

\caption{\label{fig:diagram9a} The mass reconstruction of pseudo scalar Higgs A for second process.}
\end{figure}

 In our assumed signal process, the decay of A Higgs boson occurs as $A \rightarrow ZH$. 
In accordance with that scenario, the jets which successfully passes  the cuts, A candidate are considered by identifying the combination of bjets.
 In events with possible two candidates of H, smallest value of the parameter $\Delta R $ of the combination of ZH is clearly assumed as A candidate. 
The plot for reconstruction of  pseudo scalar Higgs A for first and second process are shown in figures Figure~\ref{fig:diagram9} and Figure~\ref{fig:diagram9a} respectively.
\subsection{Mass Reconstruction of Z boson}
The jets are assumed as the light jets which do not fulfill the requirements for the declaration of b-jets. For obtaining the reconstructed invariant mass of Z boson, two of the leading jets are mainly selected which have same $\eta$ and $p_{T}$ cuts implemented on all jets. The feature, low jet multiplicity of the signal events, is used for the suppression of the Z single events. The two light jets having the highest  $p_{T}$  are fused together to create the candidate of Z-boson.\\ 
The functions which are properly fit are mainly fitted to the distributions of signal and the finalized results are indicated with the error bars. The significant peaks in the fitted profiles are near to the Higgs masses which are generated in our scenario. ROOT 6.20 \cite{root} is used for the fitting process. The Gaussian function was observed in the signal distributions of fit functions. The signal was covered by the Gaussian function. To find out the central region of a signal peak the parameter “Mean” is used which is the parameter of the fit function of Gaussian. Values of the parameter Mean are assumed as reconstructed masses of Higgs bosons which are termed as $ m_{Rec} $. For the comparison, it also indicates the generated masses which are termed as $m_{Gen}$. There is difference between the reconstructed and generated masses  due to arising uncertainties, from the algorithm of jet clustering and the mis-identification of jets, rate of mis-tagging jet, method of fitting and the selection of the fit function, errors arising in the energy and the momentum of particles, etc. The error factor may be reduced by the process of optimization of the algorithm of jet clustering, the algorithm of b-tagging and the method of fitting. Generated, reconstructed and corrected reconstructed mass of neutral scalar Higgs H1 and Pseudo scalar Higgs A is shown in Table~\ref{tab:signalcuts4} and Table~\ref{tab:signalcuts5}.
 \begin{table}[h!]
\centering
\begin{tabular}{|c|c|c|c|c||c|c|}
\hline
		
		&$m_{Gen}$ & $m_{Rec}$ &$m_{Cor. Rec.}$ \\
		\hline
		BP1 &150&140.3$\mp$0.1&	152.86$\mp$0.2   \\
		\hline
		BP2&200&	186.4$\mp$0.1&	198.96$\mp$0.2 \\
		\hline
		BP3&	250&	236.5$\mp$0.1&	249.06$\mp$0.2\\
		\hline
		BP4&300&	286.5$\mp$0.1&	299.06$\mp$0.2\\

		\hline

\end{tabular}
\caption{\label{tab:signalcuts4}The  generated, reconstructed and corrected reconstructed mass of neutral scalar Higgs H1 in first process.}
\end{table}
 \begin{table}[h!]
\centering
\begin{tabular}{|c|c|c|c|c||c|c|}
\hline
		
		&$m_{Gen}$ & $m_{Rec}$ &$m_{Cor. Rec.}$ \\
		\hline
		BP1  &250& 229.96$\mp$0.16&	243.84$\mp$0.43  \\
		\hline
		BP2&300&	277.2$\mp$0.12&	291.08$\mp$0.0.39 \\
		\hline
		BP3 &	330&	337.8$\mp$0.61&	323.92$\mp$0.0.88\\
		\hline
		BP4 &	400&	379.89$\mp$0.18&	393.77$\mp$0.45\\
		
		\hline
		
\end{tabular}
\caption{\label{tab:signalcuts5} The generated, reconstructed and corrected reconstructed mass of Pseudo scalar Higgs A in first process .}
\end{table}
From Table~\ref{tab:signalcuts4} and Table~\ref{tab:signalcuts5}, the average difference between the generated mass and reconstructed mass of $H_{1}$ and A is 12.56  and 13.88 respectively for first process, and average mass error is 0.1 and 0.27 respectively. This error is reduced by adding a same value in reconstructed mass and fill in $m_{Corr. rec.} $ column. Similarly average value of error is added in all error values and given in third column. From the Table~\ref{tab:signalcuts4} and Table~\ref{tab:signalcuts5} it is concluded that reconstructed mass is few GeV different from generated mass. Generated, reconstructed and corrected reconstructed mass of neutral scalar Higgs H and Pseudo scalar Higgs A is shown in Table~\ref{tab:signalcuts4a} and Table~\ref{tab:signalcuts5a}.
 \begin{table}[h!]
\centering
\begin{tabular}{|c|c|c|c|c||c|c|}
\hline
		
		&$m_{Gen}$ & $m_{Rec}$ &$m_{Cor. Rec.}$ \\
		\hline
		BP1 &150&139.9$\mp$0.1&	153.8$\mp$0.13   \\
		\hline
		BP2&200&	186.5$\mp$0.009&	200.4$\mp$0.039 \\
		\hline
		BP3&	250&	234.4$\mp$0.007&	248.3$\mp$0.037\\
	\hline
		BP4 &	300&	283.6$\mp$0.007&	297.5$\mp$0.0.037\\
		\hline

\end{tabular}
\caption{\label{tab:signalcuts4a}The  generated, reconstructed and corrected reconstructed mass of neutral scalar Higgs H in second process.}
\end{table}
 \begin{table}[h!]
\centering
\begin{tabular}{|c|c|c|c|c||c|c|}
\hline
		
		&$m_{Gen}$ & $m_{Rec}$ &$m_{Cor. Rec.}$ \\
		\hline
		BP1  &250& 233.9$\mp$0.1027&	250.95$\mp$0.1327  \\
		\hline
		BP2&300&	282.5$\mp$0.008&	299.95$\mp$0.038 \\
		\hline
		BP3 &	330&	312.383$\mp$0.007&	329.433$\mp$0.037\\
		\hline
		BP4 &	400&	383$\mp$0.007&	400.05$\mp$0.037\\
		\hline

\end{tabular}
\caption{\label{tab:signalcuts5a} The generated, reconstructed and corrected reconstructed mass of Pseudo scalar Higgs A in second process .}
\end{table}
From Table~\ref{tab:signalcuts4a} and Table~\ref{tab:signalcuts5a}, the average difference between the generated mass and reconstructed mass of H and A is 13.9 and 17.05 respectively for second process, and average mass error is 0.03 and 0.03 respectively. This error is reduced by adding a same value in reconstructed mass and fill in $m_{Corr. rec.} $ column. Similarly average value of error is added in all error values and given in third column. From the Table~\ref{tab:signalcuts4} and Table~\ref{tab:signalcuts5} it is concluded that reconstructed mass is few GeV different from generated mass.

\section{Signal significance}
To figure out the observability in our assumed scenario of A and H Higgs bosons, for each one of the distribution of the candidate mass, the significance of signal is computed. The numbers of candidate masses in signal and background events are counted in whole mass series. Mainly the jets are not easily detected as the b-jets due to the production of several jets in the ongoing events. In the assumed scenario,  several techniques are used to identify the jets. The associated jets are identified through the process of tagging which is aptly known as the algorithm of b-tagging. 
To identify the b-jets, the minimum distance between the b-parton and all of the generated jets is calculated. The term delta R  is helpful in finding out the b-jets by calculating the distance between the b-parton and the jets. The computation of significance of signal is totally based on the luminosity (Integrated) of 100 $f{b}^-1$, 500 $f{b}^-1$, 1000 $f{b}^-1$ and 5000 $f{b}^-1$ for all of our profiles of mass distribution.
 The Table~\ref{tab:10z} shows the signal significance values at each benchmark point, for neutral Higgs H in first process at integrated  luminosities of $100 fb^{-1}$, $500 fb^{-1}$,$1000 fb^{-1}$ and $5000 fb^{-1}$ and total efficiency. 
 \begin{table}[h!]
\centering
\begin{tabular}{|c|c|c|c|c||c|c|c|c|}
\hline
&BP1&BP2&BP3&BP4\\
\hline

 Significance S/$\sqrt{B}$ at$100 fb^{-1}$ & $25.7142 $ & $ 18.1606$ &  $2.14265$&2. 74433\\
 \hline
 
 Significance S/$\sqrt{B}$ at $500 fb^{-1}$& $57.49 $ & $ 40.60$ &  $6.22 $ &6.13 \\
 
 \hline 
 Significance S/$\sqrt{B}$ at $1000fb^{-1} $& $81.31$ & $ 57.42$ &  8.80& 8.6 \\
 \hline
Significance S/$\sqrt{B}$ at $5000 fb^{-1}$ & $181.82 $ & $ 128.415$ &  $19.69$ &19.40\\
 \hline
 $\epsilon_{total}$  at $\mathcal{L} _{int}[fb^{-1}]$= 100&	0.1541&0.168993&0.0785&0.11 \\
 \hline
 $\epsilon_{total}$  at $\mathcal{L} _{int}[fb^{-1}]$= 500,1000,5000&	0.1541&0.168993&0.080&0.11 \\

 \hline
\end{tabular}
\caption{\label{tab:10z} Values of signal significance for all benchmark points at $100,500,1000$ and $5000 fb^{-1}$ for neutral Higgs $H_{1}$ in first process.}
\end{table}
Table~\ref{tab:10za} shows the signal significance values at each benchmark point, for  Pseudo scalar Higgs A in first process at integrated  luminosities of $100 fb^{-1}$, $500 fb^{-1}$,$1000 fb^{-1}$ and $5000 fb^{-1}$ and total efficiency. 
 \begin{table}[h!]
\centering
\begin{tabular}{|c|c|c|c|c||c|c|}
\hline
&BP1&BP2&BP3&BP4\\
\hline
 Significance S/$\sqrt{B}$  at$100 fb^{-1}$ & $106.535 $ & $67.76$ &  $4.71 $ &1.21 \\
 \hline
 
 Significance S/$\sqrt{B}$ at $500 fb^{-1}$& $238.219 $ & $ 151.525$ &  $10.5533 $ &3.114\\
 
 \hline 
 Significance S/$\sqrt{B}$ at $1000fb^{-1} $& $337.163$ & $ 214.288$ &  $14.92 $&4.03  \\
 \hline
Significance S/$\sqrt{B}$ at $5000 fb^{-1}$ & $753.316 $ & $ 479.163$ &  $33.3724$& 10.1189\\
\hline
$\epsilon_{total}$ at $\mathcal{L} _{int}[fb^{-1}]$= 100,1000 &0.14&0.14&0.0569755&0.093\\
\hline
$\epsilon_{total}$ at $\mathcal{L} _{int}[fb^{-1}]$= 500,5000 &0.1484&0.1484&0.0569755&0.13\\

 \hline
\end{tabular}
\caption{\label{tab:10za} Values of signal significance for all benchmark points at $100,500,1000$ and $5000 fb^{-1}$ for Pseudo scalar Higgs A in first process.}
\end{table}

\begin{figure}[h!]
 \centering 
 \includegraphics[width=0.6\textwidth]{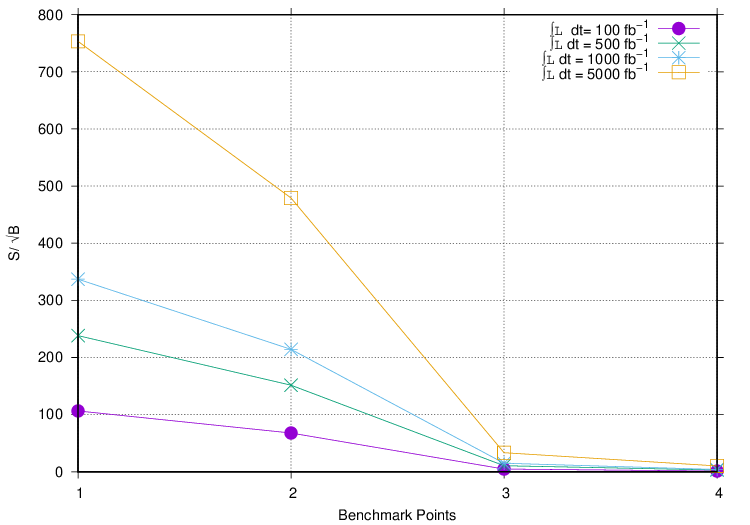}
\caption{\label{fig:30f} The Signal significance corresponding to each benchmark point at integrated luminosities of $100 fb^{-1}$, $500 fb^{-1}$, $1000 fb^{-1}$ and $5000 fb^{-1}$ for Pseudo scalar A in first process .}
 \end{figure}

\begin{figure}[h!]
 \centering 
 \includegraphics[width=0.6\textwidth]{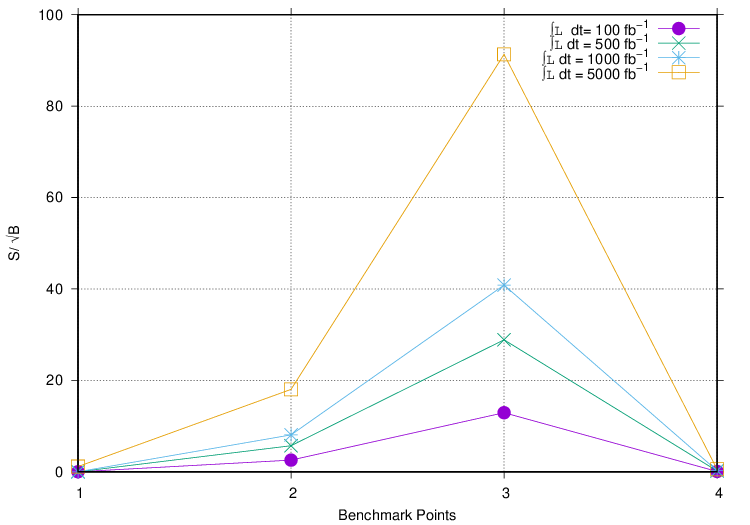}
\caption{\label{fig:30a1} The Signal significance corresponding to each benchmark point at integrated luminosities of $100 fb^{-1}$, $500 fb^{-1}$, $1000 fb^{-1}$ and $5000 fb^{-1}$ for Pseudo scalar A in second process .}
 \end{figure}

 Table~\ref{tab:10zb} shows the signal significance values at each benchmark point, for Pseudo scalar Higgs A in second process at integrated  luminosities of $100 fb^{-1}$, $500 fb^{-1}$,$1000 fb^{-1}$ and $5000 fb^{-1}$ and total efficiency for second process. 
 \begin{table}[h!]
\centering
\begin{tabular}{|c|c|c|c|c||c|}
\hline
&BP1&BP2&BP3&BP4\\
\hline
 Significance S/$\sqrt{B}$  at$100 fb^{-1}$ & $0.00135$ & $  2.55$ &  $12.914 $& 0.078\\
 \hline
 
 Significance S/$\sqrt{B}$ at $500 fb^{-1}$& 0.0030&5.70
&28.876&0.175\\
 		
 \hline 
 Significance S/$\sqrt{B}$ at $1000fb^{-1} $ &0.004269&8.066&40.8375&0.2478\\
 \hline
Significance S/$\sqrt{B}$ at $5000 fb^{-1}$ & 1.138&18.036&91.3155&0.5543\\
\hline
$\epsilon_{total}$ at	$\mathcal{L} _{int}[fb^{-1}]$= 100,500,1000 &	0.000584515&	0.153476&0.336317&0.2927 \\
 \hline
 $\epsilon_{total}$  at 	$\mathcal{L} _{int}[fb^{-1}]$ = 5000&	0.30916&	0.153476&0.336317&0.2927 \\
 \hline
\end{tabular}
\caption{\label{tab:10zb} Values of signal significance for all benchmark points at $100,500,1000$ and $5000 fb^{-1}$ for Pseudo scalar Higgs A in second process.}
\end{table}
Table~\ref{tab:10zc} shows the signal significance values at each benchmark point, for  Neutral Higgs H in second process at integrated  luminosities of $100 fb^{-1}$, $500 fb^{-1}$,$1000 fb^{-1}$ and $5000 fb^{-1}$ . 
 \begin{table}[h!]
\centering
\begin{tabular}{|c|c|c|c|c||c|c|}
\hline
&BP1&BP2&BP3&BP4\\
 Significance
 		S/$\sqrt{B}$at $100 fb^{-1}$	&0.001382& 3.099&12.60&0.07411\\
 \hline
 
 Significance S/$\sqrt{B}$	at $500 fb^{-1}$&0.00309&6.93&28.1776&0.1657\\
 
 \hline 
 Significance S/$\sqrt{B}$	at $1000 fb^{-1}$&0.004373&9.80064&39.8492&0.2343 \\
 \hline
Significance S/$\sqrt{B}$	at $5000 fb^{-1}$&1.01852&21.9149
&89.1055&0.5240\\
\hline
 		$\epsilon_{total}$ at $\mathcal{L} _{int}[fb^{-1}]$ = 100,500,1000&0.000633225&0.197228&0.3471&0.2927\\
 		
 		\hline
 		$\epsilon_{total}$ at $\mathcal{L} _{int}[fb^{-1}]$ = 5000 &0.37&0.197228&0.3471&0.2927\\
 		\hline
\end{tabular}
\caption{\label{tab:10zc} Values of signal significance for all benchmark points at $100,500,1000$ and $5000 fb^{-1}$ for Neutral  Higgs H in second process.}
\end{table}
The signal significance,  plotted verses benchmark points is shown in Figure~\ref{fig:30f} for first process and for second process it is shown in  Figure~\ref{fig:30a1}.  

\begin{figure}[h!]
 \centering 
 \includegraphics[width=0.6\textwidth]{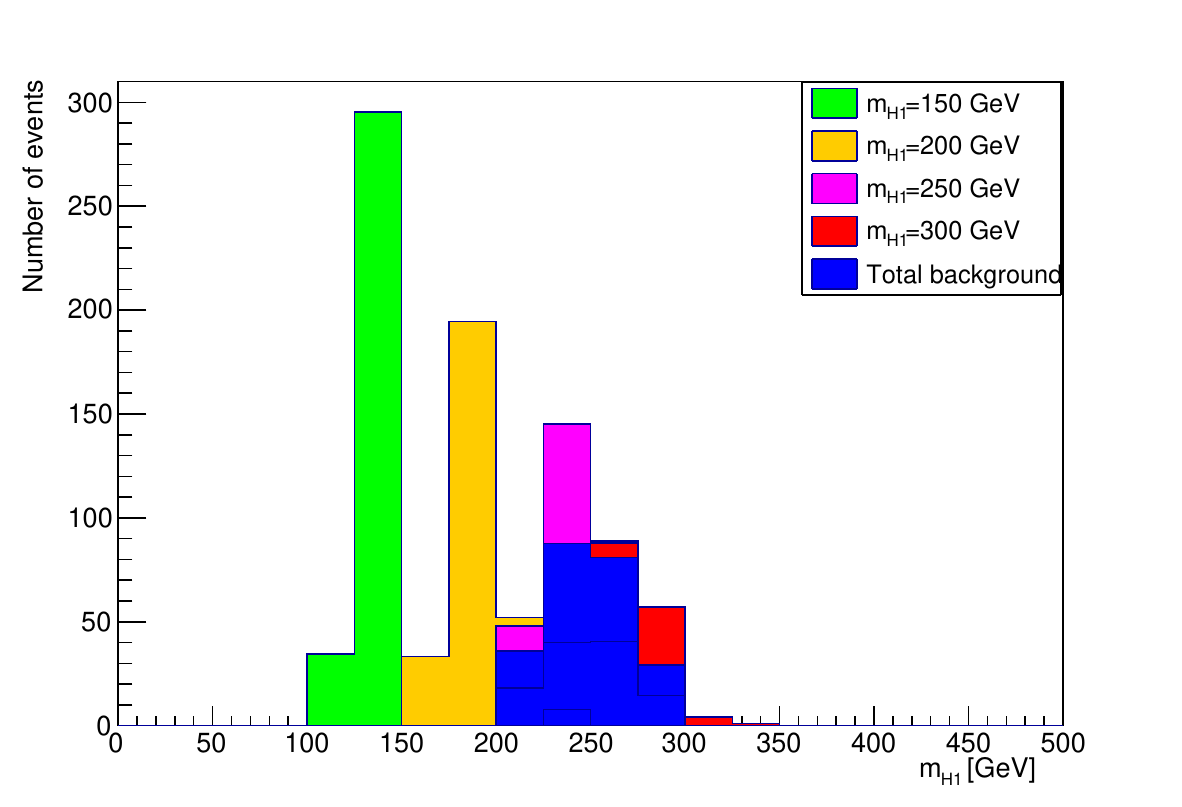}
 
\caption{\label{fig:11a} The Signal significance corresponding to each benchmark point at integrated luminosities of $500 fb^{-1}$ for neutral Higgs $ H_{1}$ for first process.}
 \end{figure}

\begin{figure}[h!]
 \centering 
 \includegraphics[width=0.6\textwidth]{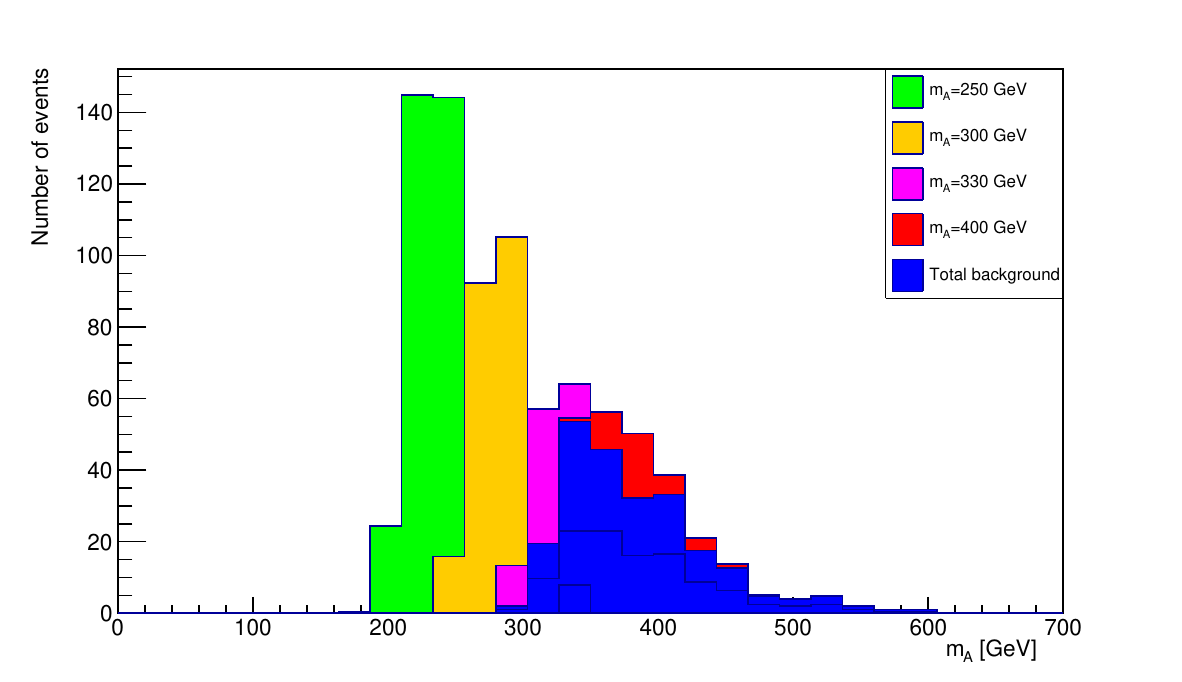}

\caption{\label{fig:31f} The Signal significance corresponding to each benchmark point at integrated luminosities of $500 fb^{-1}$ for Pseudo scalar Higgs A for first process.}
 \end{figure}

\begin{figure}[h!]
 \centering 
 \includegraphics[width=0.6\textwidth]{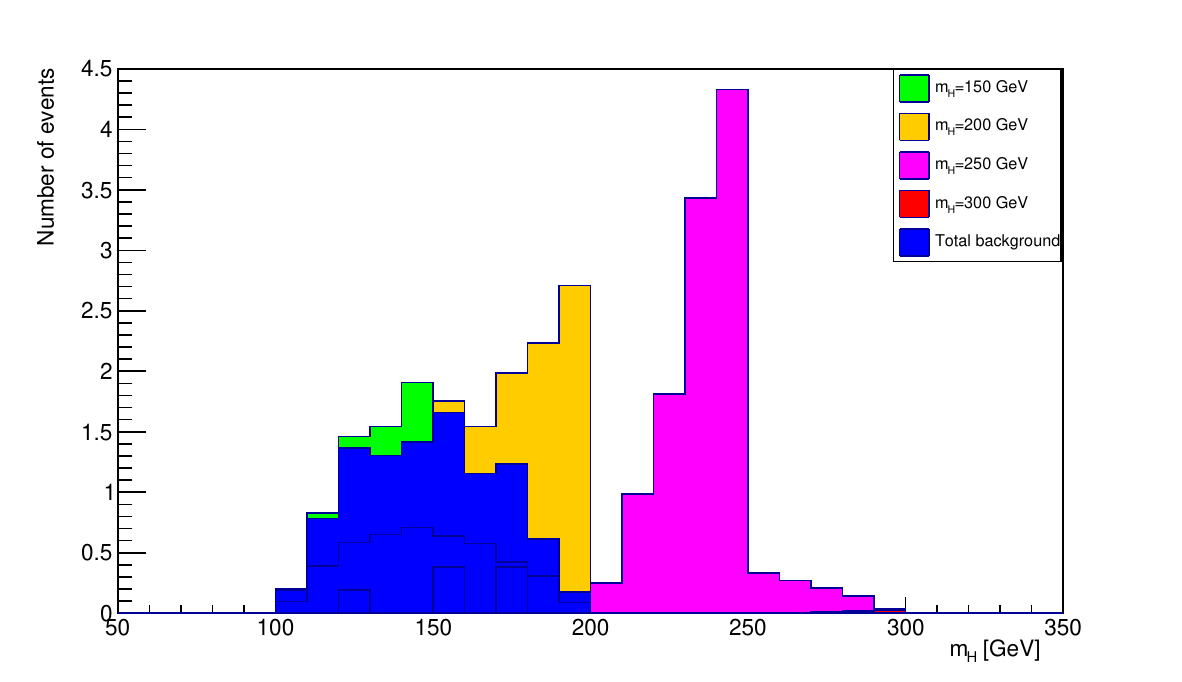}
\caption{\label{fig:30x} The Signal significance corresponding to each benchmark point at integrated luminosities of $500 fb^{-1}$ for neutral Higgs H for second process.}
 \end{figure}

\begin{figure}[h!]
 \centering 
 \includegraphics[width=0.6\textwidth]{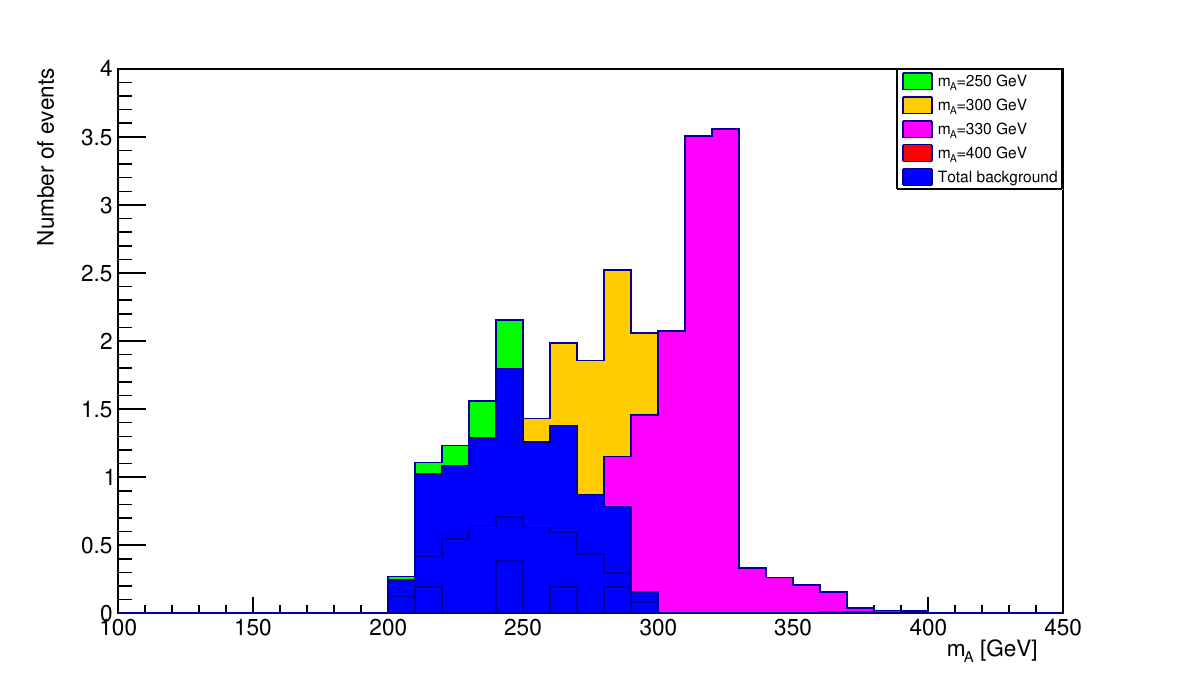}
\caption{\label{fig:31w} The Signal significance corresponding to each benchmark point at integrated luminosities of $500 fb^{-1}$ for Pseudo scalar Higgs A  for second process.}
 \end{figure}

The final results for signal significance for first and second process are shown in   Figure~\ref{fig:11a}, Figure~\ref{fig:31f}, Figure~\ref{fig:30x} and Figure~\ref{fig:31w}.

 
 
\section{Conclusion}
The  study aims to investigate the observability of Pseudo scalar Higgs A and Neutral scalar H in the framework of 2HDM type-I using lepton collider which will operate at  center of mass energy  $\sqrt{s}$=1000 GeV. The focus of study is Neutral Higgs pair production at
electron positron collider and its fully hadronic hadronic decay. The CP even Neutral Higgs decays to pair of bottom quark and the pseudoscalar Higgs decays to Z boson along with neutral heavy CP even Higgs boson. Neutral Higgs is very unstable particle which decays in no time to pair of bottom quarks. In this work, the predicted pseudo scalar (A) and neutral scalar (H) were examined using Type-I of Two Higgs doublet model(2HDM) at SM-like scenario which is  theoretical framework for this study. In 2HDM, few benchmark points (BP) in parameter space were assumed. The main chain process or the signal process is $e^{-}e^{+} \rightarrow AH \rightarrow ZHH \rightarrow jjb\bar{b}b\bar{b}$. The second process is $e^{-}e^{+} \rightarrow AH \rightarrow b \bar{b}b\bar{b}$. At low values of $ \tan \beta $, possible enhancements in the couplings of Higgs-fermion may occur. At that time the chain process gives a chance for signal to take benefit from it.
Even though,  assumed decay (hadronic) of Z boson may arise many errors and fluctuations in the final results. The  calculations due to arising uncertainties form rate of mis-tagging and the jet misidentification. Therefore  enhancement of this channel completely compensates for fluctuations and errors which arise. Few benchmark points (BP) are supposed at the $\sqrt{s}$ (center-of-mass energy) of  1000 GeV and for each BP scenario events are  generated separately. By the finalized study of data, it is
concluded that presented data analysis is the best way to observe and examine the whole scenario, which are assumed in this study. In the distributions of mass of the Higgs bosons, it can be seen that there exist significant amount of data and peaks in data of total background near the
generated masses, in the assumed luminosities (integrated). A and H Higgs bosons in all considered scenarios are observable when signal exceeds $ 5\sigma $, which is the final extracted value of signal significance in accordance with range of whole mass. Mass reconstruction was performed by the process of fitting functions to mass profiles (distributions). As a result of this process, it is concluded that in all of the assumed scenarios, the finalized reconstructed masses of Higgs bosons are in rational agreement with the
generated masses and thereby Higgs bosons (A and H) mass measurements are possible. The presented analysis is expected to work like a tool for the search of predicted neutral Higgs bosons in 2HDM. Till now simulation results, center of mass(CMS) energy and the integrated luminosities are quite promising for the observation of all the assumed scenarios.


\end{document}